\documentclass[aps,prl,superscriptaddress,twocolumn,showpacs,floatfix]{revtex4}
\usepackage{epsfig}

\newcommand{\ignore}[1]{}

\begin{document}
\title{Detecting fuzzy community structures in complex networks with a
Potts model}

\author{J\"org Reichardt}
\author{Stefan Bornholdt}
\affiliation{Interdisciplinary Center for Bioinformatics, University of
Leipzig, Kreuzstr.\ 7b, D-04103 Leipzig, Germany}
\affiliation{Institute for Theoretical Physics, University of Bremen, 
D-28359 Bremen, Germany (present address)}
\date{\today}

\begin{abstract}
A fast community detection algorithm based on a q-state Potts model is
presented. Communities in networks (groups of densely interconnected nodes
that are only loosely connected to the rest of the network) are found to
coincide with the domains of equal spin value in the minima of a modified
Potts spin glass Hamiltonian.  Comparing global and local minima of the
Hamiltonian allows for the detection of overlapping (``fuzzy'') communities
and quantifying the association of nodes to multiple communities as well as
the robustness of a community.  No prior knowledge of the number of
communities has to be assumed.
\end{abstract}
\maketitle

Finding groups of alike elements in data is of great interest in all
quantitative sciences. For multivariate data, where the objects are
characterized by a vector of attributes, a number of efficient and well
understood clustering algorithms exist \cite{Kaufman}.  They allow to find
clusters of similar objects based on a metric between the attribute
vectors.  If, however, the data is of relational form as, e.g., a network
or graph $G(V,E)$ consisting of a set $V$ of $N$ nodes and a set $E$ of $M$
links or edges connecting them and representing some relation between the
nodes, the problem of finding alike elements corresponds to discovering
communities: sets of nodes interconnected more densely among themselves
than with the rest of the network (for a recent review see ref.\
\cite{Newman}).  For any induced subgraph $g(v,e)$ of the graph $G(V,E)$
with $n$ nodes and $m$ internal edges and $m_{nN}$ edges connecting the $n$
nodes to the $N-n$ remaining nodes of the graph, this can be formalized as:
\begin{equation}
\frac{2m}{n(n-1)} > \frac{2M}{N(N-1)} > \frac{m_{nN}}{n(N-n)}.
\label{ComDef}
\end{equation}
In other words, the inner link density should be higher than the average
link density in the network which again should be higher than the outer
density of the community. As a community structure we thus define a set of
induced subgraphs $g(v,e)$ that covers $G(N,E)$ and that fulfills
(\ref{ComDef}). Note that the problem of community detection is different
from that of minimal cut graph partitioning, as for $g(v,e)$ to be a
community it is not necessary that the number of external edges is a global
minimum. Rather, it only needs to be smaller than a certain threshold that
depends on $G(V,E)$ and the size of $g(v,e)$. We see from (\ref{ComDef})
that the presence of communities is bound to the presence of
inhomogeneities in the link distribution of a graph. Furthermore, it is
understood that a community structure is not defined uniquely on a
network. Rather, a number of community structures different in size and
number of communities may exists that all fulfill the inequalities
(\ref{ComDef}). Certain nodes may belong to the same community in one
realization and may be assigned to a different community in another
realization. The differences and similarities of these realizations yield
valuable information about the robustness of a particular community
structure. Furthermore, the nodes which can be assigned into more than one
community represent an overlap of possible community structures that cannot
be interpreted as a hierarchy of communities, since the overlap may only be
partial. Here, we introduce a new algorithm that can rapidly detect a
community structure and allows for a quantitative assessment of the
individual realizations.

In this paper, we combine the early idea by Fu and Anderson for graph
bi-partitioning with a modified Ising Hamiltonian \cite{FuAnd} and the
recent Potts model clustering of multivariate data by Blatt et al.\
\cite{Blatt}. This will allow us to map the communities of a network onto
the magnetic domains in the ground state or in local minima of a suitable
Hamiltonian. For this purpose we alter a q-state Potts Hamiltonian by
adding a global constraint that forces the spins into communities according
to (\ref{ComDef}):
\begin{equation}
\mathcal{H}=-J\sum_{(i,j)\in
E}\delta_{\sigma_i,\sigma_j}+\gamma\sum_{s=1}^{q}\frac{n_s(n_s-1)}{2}.
\label{Hamiltonian}
\end{equation}
Here, $\sigma_i, i=1...N$ denote the individual spins which are allowed to
take $q$ values ${1...q}$, $n_s$ denotes the number of spins that have spin
$s$ such that $\sum_{s=1}^q n_s=N$, $J$ is the ferromagnetic interaction
strength, $\gamma$ is a positive parameter, and $\delta$ is the Kronecker
delta.  The first sum is the standard ferromagnetic Potts term for nodes
connected by an edge in the network, and is minimized by
$\mathcal{H}_{\mbox{ferr}}=-JM$. It favors a homogeneous distribution of
spins over the network. Diversity, on the other hand, is introduced by the
second term which sums up all possible pairs of spins which have equal
value. It counter-balances the first sum and increases the energy with
increasing homogeneity of the spin configuration. It represents a global
anti-ferromagnetic interaction being maximal when all nodes have the same
spin, and minimal when all possible spin values are evenly distributed over
all nodes.

The choice of $\gamma$ determines how strongly the minimum of the combined
Hamiltonian depends on the topology of the network. Consider a network of
two communities $g_1(n_1,m_1)$ and $g_2(n_2,m_2)$ with $m_{12}$ edges
connecting them. For the ground state to be composed of these two
communities, $\gamma^*$ has to obey a simple condition
\begin{equation}
\nonumber \mathcal{H}_{\mbox{homogeneous}} \geq \mathcal{H}_{\mbox{diverse}}
\end{equation}
\begin{eqnarray}
\nonumber -J(m_1+m_2+m_{12})+\gamma^{*}\frac{(n_1+n_2)(n_1+n_2-1)}{2} \geq \\
\nonumber -J(m_1+m_2) + \gamma^{*}(\frac{n_1(n_1-1)}{2}+\frac{n_2(n_2-1)}{2})
\end{eqnarray}
\begin{equation}
\gamma^{*}\geq J\frac{m_{12}}{n_1n_2}. 
\label{GammaMeaning}
\end{equation}
Comparing with (\ref{ComDef}) we see that, apart from the ferromagnetic
coupling $J$, $\gamma^*$ is just the outer link density of community 
$g_1(n_1,m_1)$. 
Thus, with the parameter $\gamma$ we enforce a ground state of the system
such that all groups of nodes with equal spin have a an outer link density
smaller than $\gamma$.
Setting $J=1$ and $\gamma$ to be the average connection probability of the
network $p=\frac{2M}{N(N-1)}$ (or $\gamma^{*}=p\langle J_{ij}\rangle$ for
weighted networks), we thus satisfy the second inequality in
(\ref{ComDef}). The first inequality in (\ref{ComDef}) is satisfied
implicitly, because high inner link densities are energetically 
favored 
by the Hamiltonian. Different local minima of the Hamiltonian then correspond
to different possible assignments of community structures.  It is
instructive to write the Hamiltonian (\ref{Hamiltonian}) in the form
\begin{equation}
\mathcal{H}=\sum_{i<j}\delta(\sigma_i,\sigma_j)(\gamma-J_{ij})
\label{HamiltonianSpin}
\end{equation}
with $J_{ij}$ as the (weighted) adjacency matrix of the graph. The ground
state structure of this spin glass Hamiltonian corresponds to the community
structure of the network. Fortunately, finding the ground state is
difficult only for random networks which usually do not exhibit any clear
community structure, and where the ambiguous community assignment
corresponds to a typical spin glass situation of multiple local energy
minima. Relevant examples of networks with non-random community structure,
however, usually correspond to Hamiltonians with prominent ground states in
large basins of attraction which makes our approach particularly
practicable.

The number of possible communities $q$ is not a critical parameter in the
algorithm: it only needs to be chosen large enough to accommodate for all
possible communities. If the number of communities is smaller than $q$, the
remaining spin states will not be populated.  However, since the runtime of
the algorithm is linear in $q$, a reasonable value should be chosen
($q<100$ was sufficient in our case).

It remains to define a measure of the statistical significance of the
communities found. Given the number of nodes in a community $n$, the number
of inner links $l_{in}$, and the number of outer links $l_{out}$ we can
calculate the expected number of possible equivalent communities
$E(n,l_{in},l_{out})$ in a random network of the same size ($N$,$M$) and
connection probability $p=\frac{2M}{N(N-1)}$:
\begin{eqnarray}
\nonumber E(n,l_{in},l_{out})=\left( \begin{array}{c} N \\ n
       \end{array}\right) \left( \begin{array}{c} \frac{n(n-1)}{2} \\
       l_{in}
       \end{array}\right)
       \left( \begin{array}{c} n(N-n) \\ l_{out} \end{array}\right)\times\\
       p^{l_{in}}(1-p)^{\frac{n(n-1)}{2}-l_{out}}p^{l_{out}}(1-p)^{n(N-n)-l_{out}}
\end{eqnarray}
If $E(n,l_{in},l_{out})$ is larger than $1$, we can expect to find such a
community in a random network of the same size, marking the border of
statistical significance.

To practically find or approximate the ground state of our system we employ
a simple Monte-Carlo heat-bath algorithm with simulated annealing
\cite{Kirkpatrick}.  Starting from a temperature with an acceptance ratio
of $>95\%$, the system is subsequently cooled down using a decrement
function for the temperature of the form $T_{k+1}=\alpha T_k$ with
$\alpha=0.99$ or similar values for the $k^{th}$ step, until it reaches a
configuration where no more than a given number of spin flips are accepted
during a certain number of sweeps over the network. In one such run, one
reaches the ground state or another low lying local minimum that
corresponds to a community structure of the network. With a set of
several runs, we are able to evaluate the robustness of the community
classification by sampling the local minima of the energy landscape of the
Hamiltonian.  The number of co-appearances of nodes in one community are then
binned in an $N\times N$ matrix. We then order the rows and columns of this
matrix according to the assignment of communities from a single simulated
annealing run. Well defined community structures then appear as blocks of
high co-appearance along the diagonal. Off-diagonal instances of high
co-appearance indicate a possible overlap between clusters.

\begin{figure}
\epsfig{file=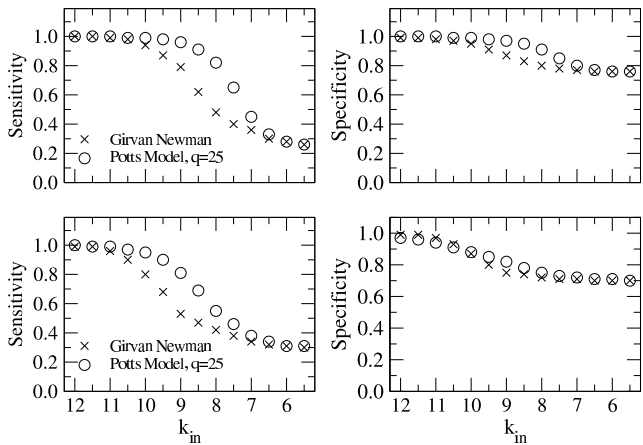, width=8.5cm}
\caption{Benchmark of the algorithm for networks with known community
structure and comparison with Girvan and Newman. Top row: 
4 communities of 32 nodes each, bottom row: 4 communities of 128, 96, 64 
and 32 nodes respectively. Symbol size corresponds to error bars.}
\label{Figure1}
\end{figure}

Let us first check our algorithm by applying it to a number of
computer-generated random test networks with known community structure as
suggested in \cite{FastGN}. Nodes are assigned to communities and are
randomly connected to members of the same community by an average of
$\langle k_{in}\rangle$ and to members of different communities by an
average of $\langle k_{out} \rangle$ links. Fixing the average degree of
all nodes to $\langle k\rangle=\langle k_{in}\rangle+\langle
k_{out}\rangle=16$, it becomes more and more difficult for any algorithm to
detect the communities as $\langle k_{in}\rangle$ decreases on the expense
of $\langle k_{out}\rangle$. Sensitivity and specificity are benchmarked
over all possible pairs of nodes. As true positive (negative) we count a
pair of nodes that is in the same (a different) community by design and is
classified accordingly by the algorithm. We tested two sets of
networks. The first is composed of four equally sized communities of 32
nodes each and the second is composed of four communities of 128, 96, 64
and 32 nodes respectively. Performance of our algorithm and, for
comparison, the one by Girvan and Newman (GN) \cite{Girvan} is shown in
Figure \ref{Figure1}.
Note the high sensitivity and specificity of our algorithm for both types of
networks. When running our algorithm without simulated annealing, but
simply relaxing the system at temperature zero from a random initial
condition it is extremely fast, yet still performs as good as the GN method. 

\begin{figure}
\epsfig{file=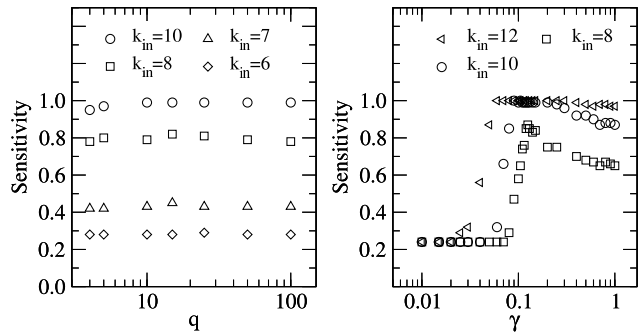, width=8.5cm}
\caption{Robustness of results for the test network with 4 communities of 
32 nodes each. Left: Sensitivity vs.\ $q$ at the end of a Monte-Carlo
optimization at different values of $\langle k_{in}\rangle$. Averaged
over 50 graphs. Right: Sensitivity for $q=25$ as a function of
$\gamma$ for different values of $\langle k_{in}\rangle$. All results averaged 
over 10 graphs.}
\label{Figure2}
\end{figure}

Figure \ref{Figure2} shows the dependence of the sensitivity of the
algorithm on $q$ in the case of the test network with equally sized
communities for four different values of $\langle k_{in} \rangle$. Note
that results do not depend on $q$. For the same type of test networks,
Figure \ref{Figure2} also shows the robustness of the sensitivity with respect
to the choice of $\gamma$. The better the communities
are defined (the larger $\langle k_{in}\rangle$), the more robust are the
results. The maxima of the curves for all values of $\langle
k_{in}\rangle$, however, coincide at $\gamma=p\simeq 0.125$ which again justifies
this choice of parameter. The same statements apply to the specificity.

\begin{figure}[t]
\epsfig{file=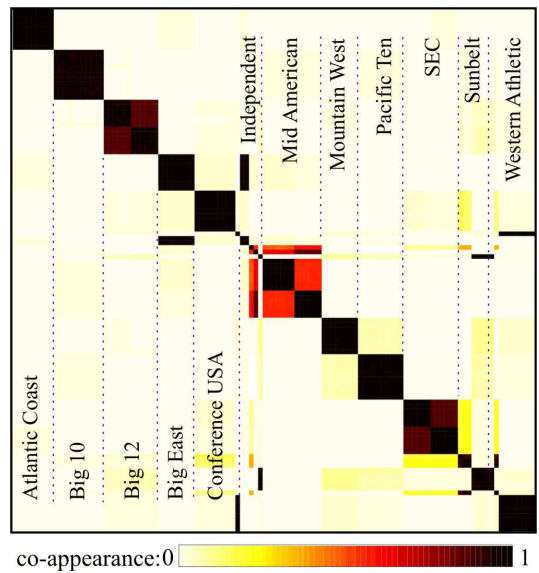, width=7.0cm}
\caption{Co-appearance matrix for the football
  network. $0.1p\leq\gamma\leq p$, matrix ordering taken from assignment
  of teams into conferences according to game schedule.}
\label{Figure3}
\end{figure}

One real world example with known community structure is the College
Football network from ref.\ \cite{Girvan}.  It represents the game schedule
of the 2000 season of Division I of the US college football league. The
nodes in the network represent the 115 teams, while the links represent 613
different games played in the course of the year. The community structure
of this network arises from the grouping into conferences of 8-12 teams
each. On average, each team has 7 matches with members of its own
conference and another 4 matches with members of different conferences. We
perform a parameter variation in $\gamma$ at ten values between
$0.1p\leq\gamma\leq p$. At each value of $\gamma$ we relax the system 50
times from a randomly assigned initial configuration at $T=0$ using
$q=50$. Figure \ref{Figure3} shows the resulting $115\times 115$
co-appearance matrix, normalized and color coded. The ordering of the
matrix corresponds to the assignment of the teams into conferences
according to the game schedule. The dashed blue lines indicate this. Apart
from regaining almost exactly the known community structure, our algorithm
is also able to detect inhomogeneities in the distribution of intra- and
inter-conference games. For instance, we see a large overlap of the Pacific
Ten and Mountain West conference and also a possible subdivision of the Mid
American conference into two sub-conferences, one of which contains Ball
State, Toledo, Central, Eastern, Northern and Western Michigan. This is due
to the fact, that geographically close teams are more likely to play
against each other as already pointed out in ref.\ \cite{Girvan}.

Finally we consider a large real world example with only partially known
community structure, a large protein folding network compiled by Rao and
Caflisch \cite{Rao}. This network represents the conformation space of a 20
amino acids peptide sampled by molecular dynamics at the melting
temperature. $5\times10^5$ subsequent conformational snapshots were taken
at time intervals of 20ps, resulting in 132168 different configurations
sampled and 228972 observed transitions between two different
conformations. These represent a network of conformations, where a link
indicates that two conformations follow each other in time. Analysis of
this network yields valuable information about the free energy landscape of
the folding Hamiltonian without the need of projecting it onto arbitrarily
chosen coordinates.  Applying the algorithm to the complete unweighted
network using $q=50$ and $\gamma=p$ yields a largest community of 16,000
nodes, correctly corresponding to the folded state (FS). The statistical
weight of the nodes in this community was found to be $55\%$ of the total
weight which confirms the expectation of the folded and denatured state
being equally populated at the melting temperature.
\begin{figure}[t]
\epsfig{file=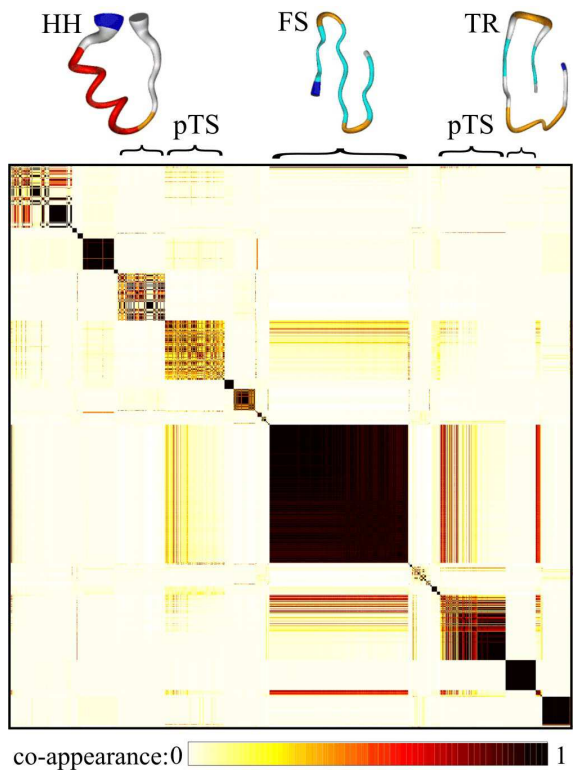, width=7.5cm}
\caption{Co-appearance matrix for the reduced version of the protein
  folding network. $0.1p\leq\gamma\leq p$, matrix ordering taken from a
  simulated annealing run of the full network.}
\label{Figure4}
\end{figure}
The characteristic conformations of the denatured state, the high enthalpy,
high entropy conformations, such as the helical conformations (HH), as well
as low entropy conformations such as the curl like trap (TR) are also
recognized as communities.  Again, $\gamma$ is varied between
$0.1p\leq\gamma\leq p$ and $T=0$ with 50 repetitions at each value of
$\gamma$ and $q=50$. For this, we used the reduced version of the folding
network as in \cite{Rao} that contains only nodes which are visited 20
times or more in the course of the simulation, resulting in 1287 nodes and
23948 links. Figure \ref{Figure4} shows the resulting $1287\times 1287$
nodes co-appearance matrix. The rows and columns are ordered with respect
to one single simulated annealing run at $\gamma=p$.  Thus, we see how well
the ground state is approximated by the local minima and how robust the
assignment into communities is with respect to $\gamma$.  Again we find a
clear characterization of the FS and TR communities.  The helical
conformations (HH), however, do not occur in one community for all values
of $\gamma$ which indicates many different possible assignments into
communities and is an indication of their high entropy nature. Furthermore,
a number of putative transition states (pTS) could be assigned, that
mediate the folding from certain denatured configurations into the folded
state.

In conclusion, we discuss a new algorithm for community detection in
complex networks based on a modified q-state Potts model.  Communities
appear as domains of equal spin value near the ground state of the system.
which is approximated through Monte-Carlo optimization. Only local
information is used to update the spins which makes parallelization of the
algorithm straightforward and allows the application to very large
networks. On both, computer-generated and real world networks as studied
here the algorithm performs fast, often considerably faster than current
state-of-the-art algorithms.  Without using prior knowledge it
automatically detects the number of communities as the number of occupied
spin states. As the algorithm is non-deterministic and non-hierarchical, it
allows for the quantification of both, the stability of the communities, as
well as the affiliation of a node to more than one community (``fuzzy
communities'').

The authors 
thank K.\ Klemm for many inspiring discussions and 
 A.\ Caflisch, M.\ Newman, and F.\ Rao for the supply of network data and
helpful comments.

\bibliography{BibTex_Citations}

\begin{thebibliography}{8}
\expandafter\ifx\csname natexlab\endcsname\relax\def\natexlab#1{#1}\fi
\expandafter\ifx\csname bibnamefont\endcsname\relax
  \def\bibnamefont#1{#1}\fi
\expandafter\ifx\csname bibfnamefont\endcsname\relax
  \def\bibfnamefont#1{#1}\fi
\expandafter\ifx\csname citenamefont\endcsname\relax
  \def\citenamefont#1{#1}\fi
\expandafter\ifx\csname url\endcsname\relax
  \def\url#1{\texttt{#1}}\fi
\expandafter\ifx\csname urlprefix\endcsname\relax\def\urlprefix{URL }\fi
\providecommand{\bibinfo}[2]{#2}
\providecommand{\eprint}[2][]{\url{#2}}

\bibitem[{\citenamefont{Kaufman and Rousseeuw}(1990)}]{Kaufman}
\bibinfo{author}{\bibfnamefont{L.}~\bibnamefont{Kaufman}} \bibnamefont{and}
  \bibinfo{author}{\bibfnamefont{P.}~\bibnamefont{Rousseeuw}},
  \emph{\bibinfo{title}{Finding Groups in Data: an introduction to cluster
  analysis}} (\bibinfo{publisher}{Wiley-Interscience}, \bibinfo{year}{1990}).

\bibitem[{\citenamefont{Newman}(2004{\natexlab{a}})}]{Newman}
\bibinfo{author}{\bibfnamefont{M.~E.~J.} \bibnamefont{Newman}},
  \bibinfo{journal}{Eur. Phys. J. B} \textbf{\bibinfo{volume}{38}}
  (\bibinfo{year}{2004}{\natexlab{a}}).

\bibitem[{\citenamefont{Fu and Anderson}(1986)}]{FuAnd}
\bibinfo{author}{\bibfnamefont{Y.}~\bibnamefont{Fu}} \bibnamefont{and}
  \bibinfo{author}{\bibfnamefont{P.~W.} \bibnamefont{Anderson}},
  \bibinfo{journal}{J. Phys. A: Math. Gen.} \textbf{\bibinfo{volume}{19}},
  \bibinfo{pages}{1605} (\bibinfo{year}{1986}).

\bibitem[{\citenamefont{Blatt et~al.}(1996)\citenamefont{Blatt, Wiseman, and
  Domany}}]{Blatt}
\bibinfo{author}{\bibfnamefont{M.}~\bibnamefont{Blatt}},
  \bibinfo{author}{\bibfnamefont{S.}~\bibnamefont{Wiseman}}, \bibnamefont{and}
  \bibinfo{author}{\bibfnamefont{E.}~\bibnamefont{Domany}},
  \bibinfo{journal}{Phys. Rev. Lett.} \textbf{\bibinfo{volume}{76}}
  (\bibinfo{year}{1996}).

\bibitem[{\citenamefont{Kirkpatrick et~al.}(1983)\citenamefont{Kirkpatrick,
  Jr., and Vecchi}}]{Kirkpatrick}
\bibinfo{author}{\bibfnamefont{S.}~\bibnamefont{Kirkpatrick}},
  \bibinfo{author}{\bibfnamefont{C.~G.} \bibnamefont{Jr.}}, \bibnamefont{and}
  \bibinfo{author}{\bibfnamefont{M.}~\bibnamefont{Vecchi}},
  \bibinfo{journal}{Science} \textbf{\bibinfo{volume}{220}},
  \bibinfo{pages}{671} (\bibinfo{year}{1983}).

\bibitem[{\citenamefont{Newman}(2004{\natexlab{b}})}]{FastGN}
\bibinfo{author}{\bibfnamefont{M.}~\bibnamefont{Newman}},
  \bibinfo{journal}{Phys. Rev. E.} \textbf{\bibinfo{volume}{69}},
  \bibinfo{pages}{066133} (\bibinfo{year}{2004}{\natexlab{b}}).

\bibitem[{\citenamefont{Newman and Girvan}(2003)}]{Girvan}
\bibinfo{author}{\bibfnamefont{M.}~\bibnamefont{Newman}} \bibnamefont{and}
  \bibinfo{author}{\bibfnamefont{M.}~\bibnamefont{Girvan}},
  \bibinfo{journal}{Proc. Natl. Acad. Sci.} \textbf{\bibinfo{volume}{99}},
  \bibinfo{pages}{7821} (\bibinfo{year}{2003}).

\bibitem[{\citenamefont{Rao and Caflisch}(2004)}]{Rao}
\bibinfo{author}{\bibfnamefont{F.}~\bibnamefont{Rao}} \bibnamefont{and}
  \bibinfo{author}{\bibfnamefont{A.}~\bibnamefont{Caflisch}},
  \bibinfo{journal}{J. Mol. Bio.}  (\bibinfo{year}{2004}).

\end{thebibliography}

\end{document}